\documentclass[times, 10pt,twocolumn]{article} 
\usepackage{latex8}
\usepackage{times}

\usepackage{graphicx}
\usepackage{multirow}
\usepackage{booktabs}

\begin{document}

\title{
Improving Malware Detection Accuracy by Extracting Icon Information
}

\author{Pedro Silva\\
University of California, Irvine\\
Department of Computer Science\\
Irvine, CA 92697, USA\\
pedro.silva@uci.edu\\
\and
Sepehr Akhavan Masouleh \& Li Li\\
Cylance Inc.\\
Department of Research and Intelligence\\
Irvine, CA, 92612, USA\\ 
\{sakhavan-masouleh,lli\}@cylance.com\\
}

\maketitle
\thispagestyle{empty}

\begin{abstract}
Detecting PE malware files is now commonly approached using statistical and machine learning models. While these models commonly use features extracted from the structure of PE files, we propose that icons from these files can also help better predict malware. We propose an innovative machine learning approach to extract information from icons. Our proposed approach consists of two steps: 1) extracting icon features using summary statics, histogram of gradients (HOG), and a convolutional autoencoder, 2) clustering icons based on the extracted icon features. Using publicly available data and by using machine learning experiments, we show our proposed icon clusters significantly boost the efficacy of malware prediction models. In particular, our experiments show an average accuracy increase of 10\% when icon clusters are used in the prediction model. 
\end{abstract}

\section{INTRODUCTION}

Machine learning is becoming one of the more prestigious fields in computer science due to its rapid growth and record-shattering fame. In the past ten years only it has been integrated into several areas of study, such as physics~\cite{li2016pure, brockherde2016passing, li2016understanding}, chemistry~\cite{46117,46010}, 
face recognition~\cite{shao2015face}, cloud computing~\cite{bhimani2017fim,bhimani2017accelerating}, network congestion control~\cite{li2016learning,li2017dynamic}, computer vision~\cite{ponce2016chalearn,chen2016overcoming}, ocean engineering~\cite{wang2016application,wang2013new} etc., as a central or one of many pieces in the process of decision-making. The field of cybersecurity is not likely to be different~\cite{wojnowicz2016suspiciously}. Machine learning has produced a major paradigm shift in the field, being at the center of some innovative techniques for detecting and preventing infections by malicious software (malware). Most malware intend to somehow cause damage by executing malicious code on an infected machine. They do so by disguising themselves as good files. One of the most common such file is the Portable Executable (PE) file. The PE format is a standard file format for Windows executables, DLLs, object code, etc.

PE files contain several sections with information on how to map the file into memory. The files usually also come with one or more associated icons, embedded within them, and while many approaches ignore icons when performing malware analysis, we believe that information can be used to aid in the detection of malware. There are many ways this task can be achieved. One could, for example, simply train a model on pixel values. Experience and practical results show that this approach doesn't yield the best results. Most of the icons used in malware have slight blurriness, or color shifting, which is done with the exact purpose of defeating a naive approach like the one mentioned above. So, we need an approach that is robust against those subtle changes and can still provide useful information to the classifiers.

The objective of this work is to ultimately use that information to classify a PE file as good or bad. Using solely the icon to perform this task is also a bad idea. It is not uncommon for malware to use the same icon as known good files, such as a Microsoft Word document icon or an Adobe Reader PDF file icon. If we used the icon information only, we would have a dataset with conflicting labels, which would directly interfere with the performance of our classifiers. On the other hand, when combined with other features, such as size, source, content, among others, the icon information increases the accuracy of classifiers built to separate good and bad files. In the following sections, we describe in detail how we performed feature extraction and how we used that information on the classifiers.

\section{Feature Extraction} \label{feature_extraction}

The simplest way to use features from icons or images, in general, is to use raw pixel values from the three color channels -- red, green, and blue (RGB). The main problem with this approach is that it is very susceptible to noise in the images and it does not provide sufficient information about the file. This is one of the reasons why so many icons used by malware have some perturbation in the color channels. From occluded parts to blurriness in the image, and sometimes even in the form of slight increase or decrease in the RGB channel values throughout the image, which would be imperceptible to the naked eye but can cause a classifier to incorrectly classify malware as benign files \cite{Szegedy:2013vw}. These are done mainly to avoid direct matching used by some systems. In order to harvest the knowledge of these icons but at the same time circumvent these issues we decided to build three sets of features that have a great deal of information about the icon file while staying resilient against these kind of perturbations. 

\subsection{Manually Created (MC) Features}

Despite the issues that including RGB values may cause there is definitively valuable information present in these values, and we must harvest their potential somehow. To do so, we used a different approach: rather than having only pixel values, we use means and standard deviations over different sections of an image to preserve that information without being too affected by slight color variations.

\begin{figure}[thpb]
      \centering
      \includegraphics[scale=0.5]{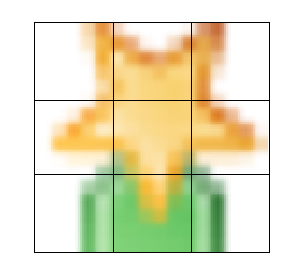}
      \caption{Icon file split in nine regions used to extract features.}
      \label{fig:iconGrid}
   \end{figure}

To achieve this, we use the mean and standard deviation (std) of the pixel values for the whole image across all the channels (2 features), then we take the mean and std of the different RGB channels (6 features). Lastly, we split the original image into nine different sections, like in Figure~\ref{fig:iconGrid}, and compute the mean and std of the pixel values on each one of the different sections across all the channels (18 features). This method produces a total of 26 MC features. 

The particular choice of splitting the image into nine regions is empirical to the problem we are trying to solve. These images are icons of files, so most of these are very small figures, usually $32\times 32$ pixel images. Increasing this $3\times 3$ grid to $4\times 4$ or $5\times 5$ brings us back to the problem we are trying to avoid by using these features: to weaken the effect of small variations in the image since the regions would be too small. Decreasing the grid to $2\times 2$, on the other hand, makes little sense since we would be losing too much information about the image and would, in fact, be creating only eight features, which would also potentially be very similar to our first 2 MC features.

\subsection{Histogram of Oriented Gradients (HOG) Features}

HOG features have been created with the idea to detect object shapes, like a hand or a person, regardless of colors~\cite{MERL_TR94-03}. The idea of using these features is to keep information despite the color fuzziness that can be seen in malware icons. HOG features capture contour, silhouette, and some texture information while providing further resistance to illumination and color variations. A small window slides over the image and computes the gradient of the image within the window~\cite{scikit-image}.

One further step that had to be done here was to ensure that the number of returned features from the HOG was always the same. Otherwise, this would affect the models during the next phase of the process, since we would not be able to guarantee all icons would produce the same number of features. After experimentation, we decided to have the HOG-parsed image be of size $24\times 24$\footnote{Most icons are either $16\times 16$, $24\times 24$, or $32\times 32$. If we decided to use $16\times 16$ as the default size we would be waisting too much information from the larger icons, and if we used $32\times 32$ we would be doubling the size of the small ones, leading to the opposite problem: fabricating too much information to perform the rescaling.}, which means we had a total of 576 features at the end. To get this, we used cell sizes of $1/8$-th of the image in both axes and resized the image whenever necessary.

\subsection{Autoencoder (AE) Features}

Up until now, both feature sets we are using involve feature engineering: thinking of ways we can analyze an image and extract meaningful information from it. Another approach to this task is rather than creating the features yourself, letting a neural network create those features. We incorporated this behavior into our model by using features generated by a convolutional autoencoder neural network, which is a neural network that models the input to itself, compressing it and decompressing it in the process~\cite{Bengio:2009kb}. In order to decompress the information, the network has to learn what are the most important features that it has to keep for each image to make it possible to recreate them with a certain accuracy. We trained the AE with a large set of icons to make it more robust and a good generalizer~\cite{He:2015tt}.

At the end of our compression step, we have a total of 512 AE features for each image. That brings us to a total of 26 (MC features) + 576 (HOG features) + 512 (AE features) = 1114 features total. We could stop the project right here and just use those features on the classifier and let it figure it out what is more useful, but we can still gather more information and reduce the number of features we would introduce to the model. One could, for example, use dimensionality reduction methods, such as random projections, principal component analysis, t-SNE, etc., to reduce the number of features from 1114 to 100, while still retaining valuable information. We wanted to take this to an extreme so we decided to tackle the problem under a different light. 

We clustered the icons and use their cluster ids as the variable the classifier would use. Similar icons tend to be close to each other even in a high dimensional space, so a cluster id is informative of the type of icons that are within the same region, thus within the same cluster. One could also perform a dimensionality reduction before the clustering, but we have not done that in this work. So, in the end, rather than using 1114 features we can use the cluster id as mentioned above in the classifier (explained in detail in sections \ref{clustering} and \ref{new_sample}).

\section{Clustering} \label{clustering}

We tested several clustering algorithms~\cite{scikit-learn,McInnes2017} for this task, namely k-means, mean shift, affinity propagation, density-based spatial clustering of applications with noise (DBSCAN), hierarchical DBSCAN (HDBSCAN), and also different hierarchical clustering techniques, like average, complete and single linkage. The ones that had more promising results were the two density-based methods: DBSCAN and HDBSCAN. 

They also have an attractive property for our purposes: they can detect outliers on the dataset. HDBSCAN outperformed DBSCAN in the quality of clusters, which were really tight and well separated. We used a silhouette score to measure their quality. An advantage of HDBSCAN is also that we do not have to provide an expected number of clusters nor a radius to look for them, like we have to for k-means or DBSCAN~\cite{Campello:2013kv}.

While HDBSCAN did an excellent job of finding the densest clusters, it was still having issues labeling a significant portion of our dataset as outliers. So to capture HDBSCAN's good properties we decided to use it and then use another clustering algorithm on the \textit{outlier} set. Basically what that means is that we are removing the super dense areas where extremely similar icons fall together and then clustering the remaining icons with another algorithm. We used a k-means algorithm to cluster the outliers.

\section{New Sample Classification} \label{new_sample}

Given a new icon, the goal is to turn it into two features: cluster id and outlier flag. The whole process is as follows:

\begin{enumerate}

\item Transform the icon into features. This is the process described in section \ref{feature_extraction}.

\item Get a cluster prediction. This is a non-trivial problem with HDBSCAN. The algorithm does not provide a prediction function. So, to perform this step, we need to make use of a classifier. A feed-forward neural network could be used in this step, but we decided to use a k-nearest neighbors (KNN) model for its simplicity and accuracy. We perform the KNN model fit using the HDBSCAN labels only, so when a new sample comes in, the model looks to its k-nearest neighbors' labels and decide by majority vote what cluster id to assign to the new sample. If a label other than -1 is assigned to the new sample, then we return said cluster id and \textit{false} for outlier detection. Getting a label of $-1$ means that HDBSCAN was likely to call this sample an outlier, which means we have to perform the prediction with the k-means model that we trained on the outlier data. If the label $-1$ is assigned to the new sample, then we set \textit{true} for outlier detection and run a subsequent prediction with k-means to get the cluster id.
\end{enumerate}

\section{EXPERIMENTS}

\begin{figure}[!h]
\centering
\includegraphics[width=1.0\columnwidth]{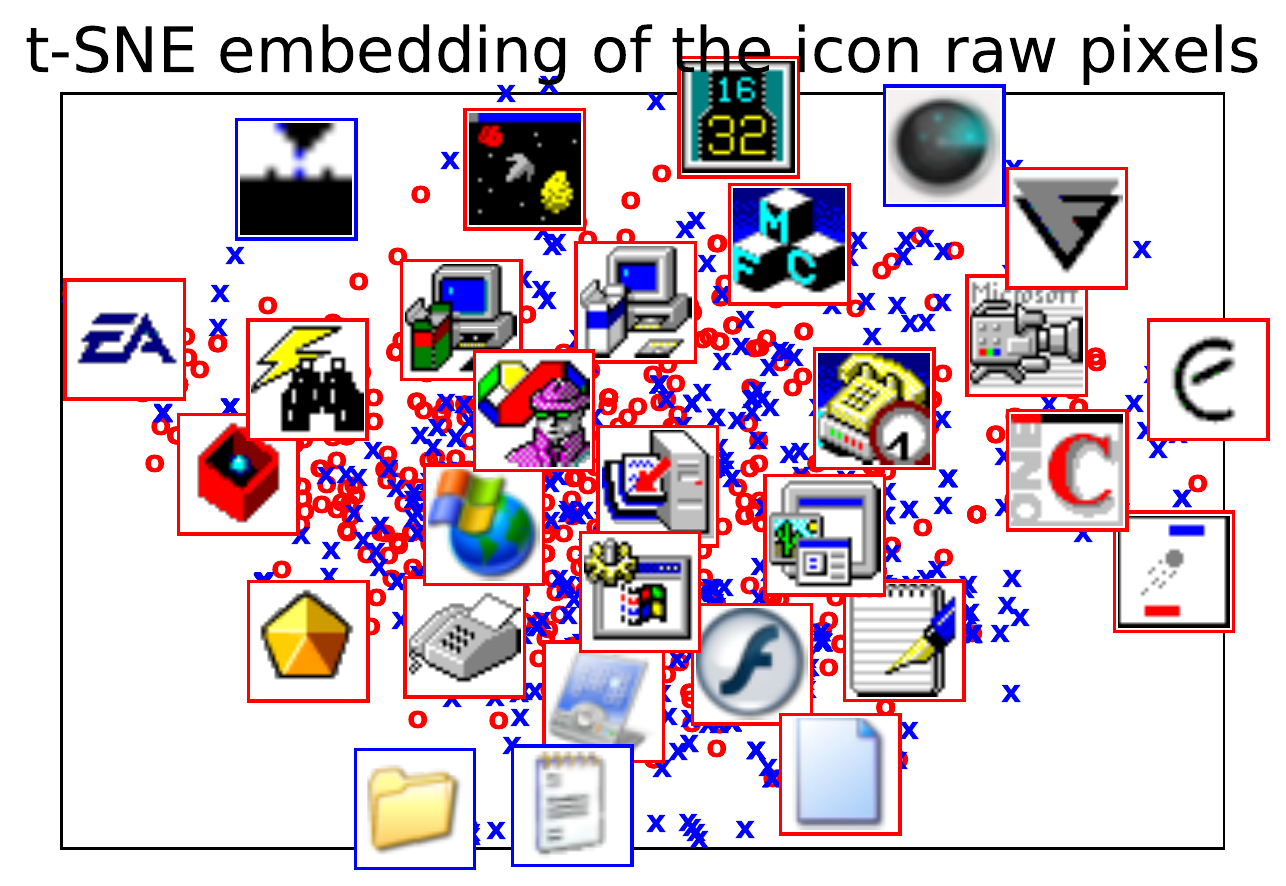}
\caption{Visualization of the PE icons used in our experiments after dimension reduction using t-SNE. Icons corresponding to benign PEs are marked with a red circle and icons corresponding to malware PEs are marked with a blue cross. Some icon images are also displayed at their locations in the t-SNE plot. Red frame indicates benign file and blue frame indicates malware.}
\label{tsne}
\end{figure}

In order to test the efficacy of our proposed method in terms of enhancement in malware prediction, we use a balanced sample of publicly available PE files obtained from virustotal.com with 1,138 benign and 1,138 malware files. In order to visualize the icons we use in the experiment (Figure~\ref{tsne}), we use t-SNE~\cite{Maaten:2008tm} on the raw icon pixels. Despite the fact that  Figure~\ref{tsne} shows malware and benign icons are well mixed, yet we will show in this section that our approach is capable of using the information in the icons to better predict detect malware. 

\begin{figure}[!b]
\centering
\includegraphics[width=0.5\textwidth]{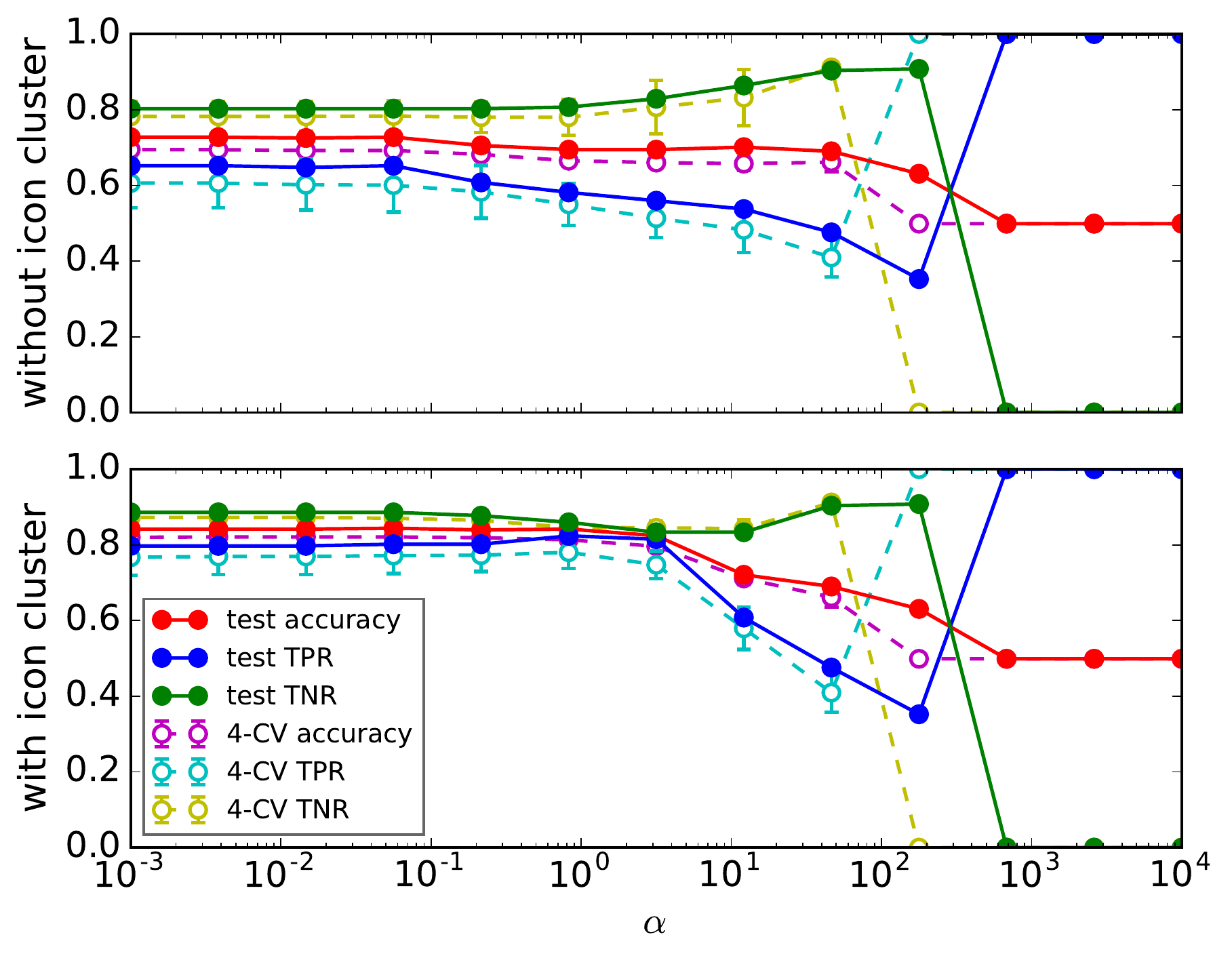}
\caption{Results from the optimization process of the regularization parameter, $\alpha$, using stratified 4-fold cross validation for the lasso logistic regression model.}
\label{cvplot}
\end{figure}

\begin{figure}[!h]
\centering
\includegraphics[width=0.75\columnwidth]{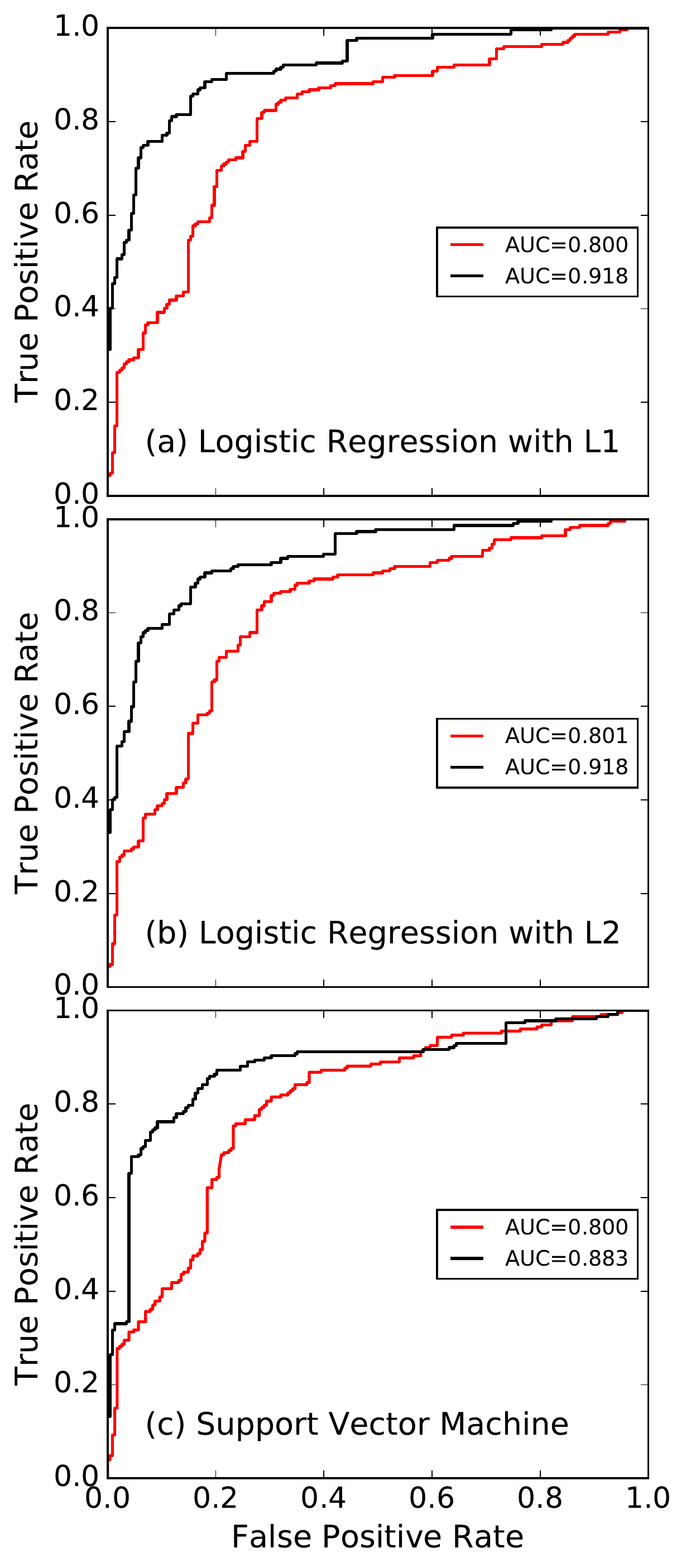}
\caption{ROC curve for the candidate models. The red line corresponds to a model using only the PEfile features. The black line corresponds to a model using the icon cluster ID feature in addition to the PEfile features.}
\label{rocFig}
\end{figure}

Using our proposed method, we initially generate icon features (MC, HOG, and AE features) and then we cluster icons. Further, per each PE file and by using the publicly available python package PEfile\footnote{ PEfile is a multi-platform Python module to parse and work with PE files. It's available at https://github.com/erocarrera/pefile}, we generate ``entropy", ``Misc\_VirtualSize", and ``SizeOfRawData" features from the three sections of ``.text", ``.data", and ``.rsrc", we shall refer to these features as \textit{PEfile} features. In order to test the effectiveness of icon clusters generated using our proposed method in better detecting malware, we build three prediction models: 1) Lasso Logistic Regression (L1), 2) Ridge Logistic Regression (L2), and 3) Linear Support Vector Machine (SVM). Each model is then fit once using only the PEfile features and another time with both PEfile features and one-hot encoded icon cluster feature\footnote{code for reproducing the experiment results is available at https://github.com/CylanceSPEAR/improving-malware-detection-accuracy-by-extracting-icon-information}.

In order to better estimate out-of-sample accuracy of the models, the original data is randomly split into train data (80\% of the data) and test data (20\% of the data). The division of the data into train and test is being done using a stratified sampling method which guarantees balanced labels in the generated train and test data. Test data remain untouched during the model fitting process and are solely used for the final out-of-sample accuracy evaluation of the model. To avoid overfitting, all of our proposed models are regularized (either L1 or L2). Regularization parameters are tuned using a stratified 4-fold cross-validation process. As an example, Figure~\ref{cvplot} shows results from the optimization process of the regularization parameter of the lasso logistic regression model using stratified 4-fold cross-validation. 

\begin{table*}
\centering
  \begin{tabular}{ccc|cccccc|cccc}
  \toprule
   \multirow{2}{*}[-0.3em]{MODEL} &  \multirow{2}{*}[-0.3em]{ICON} &  \multirow{2}{*}[-0.3em]{$\alpha$}  & \multicolumn{6}{c|}{CROSS-VALIDATION} & \multicolumn{4}{c}{TEST}\\

     & & & accuracy & std & TPR & std & TNR & std & accuracy & TPR & TNR & AUC\\
    \midrule

    \multirow{2}{*}[-0.1em]{LR L1} & no & $3.83\times 10^{-3}$ & 0.694 & 0.016 & 0.606 & 0.066 & 0.782 & 0.039 & 0.727 & 0.652 & 0.803 & 0.800 \\
     & yes & $5.62\times 10^{-2}$ & 0.821 & 0.022 & 0.772 & 0.047 & 0.870 & 0.005 & 0.844 & 0.802 & 0.886 & 0.918 \\
	
    \midrule

    \multirow{2}{*}[-0.1em]{LR L2} & no & $3.16\times 10^{-4}$ & 0.694 & 0.015 & 0.606 & 0.065 & 0.782 & 0.035 & 0.730 & 0.656 & 0.803 & 0.801 \\
     & yes & $1.00\times 10^{-5}$ & 0.823 & 0.020 & 0.769 & 0.048 & 0.876 & 0.010 & 0.842 & 0.797 & 0.886 & 0.918 \\

	\midrule

    \multirow{2}{*}[-0.1em]{SVM} & no & $3.83\times 10^{-3}$ & 0.692 & 0.021 & 0.593 & 0.098 & 0.791 & 0.059 & 0.741 & 0.692 & 0.789 & 0.800 \\
        & yes & $3.83\times 10^{-3}$ & 0.829 & 0.020 & 0.772 & 0.039 & 0.887 & 0.019 & 0.826 & 0.780 & 0.873 & 0.883 \\
   \bottomrule
\end{tabular}

\caption{Model performance results for the three candidate models of lasso logistic regression (LR L1), ridge logistic regression (LR L2), and linear SVM, each under the two scenarios of 1) icon cluster ID included as a feature in the model and 2) no icon cluster ID feature used in the model. The table includes the optimized regularization parameter, $\alpha$, cross-validation's accuracy, true positive rate, true negative rates, and their corresponding standard deviations. The table also includes accuracy, true positive and true negative rates for the models under the test data.}\label{model_performance}
\end{table*}

Table~\ref{model_performance} shows the results of fitting the three models both with icon cluster feature and without icon cluster feature . The table shows the regularization parameter, $\alpha$, that is optimized using a stratified 4-fold cross-validation. The table also includes measures of accuracy at the optimized $\alpha$ value: cross-validation accuracy and its standard error, cross-validation true positive rate and the standard error, and cross-validation true negative rate and its standard deviation. Finally, the fitted models are tested using the test data, and we report test accuracy, test true positive rate, test true negative rate, and the area under the curve. 

As one can see, adding the icon cluster feature to the feature set has consistently boosted accuracy and area under the curve across all three models. This is a clear indication that our proposed icon clustering technique can help improve malware detection in PE files. Figure~\ref{rocFig} also supports this conclusion. In this figure, we show the ROC curves associated with our three candidate models. The black curve corresponds to the model with icon cluster feature and the red curve corresponds to the model without the icon cluster feature.

\section{CONCLUSIONS}
In this paper, we proposed a new approach to incorporate information from icons of PE files into prediction models in order to better detect malware. Rather than using the raw icon pixel values, we proposed extracting features using a combination of manually created features, a histogram of gradients, and autoencoder-generated features, which led to 1,114 features. Using the extracted features, we cluster icons. This process is synonymous to reducing the extracted 1,114 features to 1 feature\footnote{Alternatively, one may also use the boolean outlier flag discussed in section \ref{new_sample} as a feature, but we decided not to use it for this paper.}, and yet still retains meaningful information about the image.

Using publicly available data, we ran experiments testing the effectiveness of our proposed icon cluster on better predicting malware. Our experiments showed a significantly higher area under the curve of the ROC plot (Figure~\ref{rocFig}) as well as an average increase of 10\% in accuracy of predicting malware when we use our proposed icon clusters inside the prediction model.  Table~\ref{model_performance} also adds a compelling argument in favor of our proposed method where we show that not only the accuracy but also the true positive and true negative rates have increased in the models with the icon cluster. This work has shown that PE icons contain useful knowledge when performing malware detection. This paper, along with many others in the field, once again shows that malware leave traces in unexpected places~\cite{wojnowicz2016suspiciously} where discovery of those hidden codes can improve the accuracy of malware prediction models.

\bibliographystyle{latex8}
\bibliography{ref}
\end{document}